\documentclass[10pt,draft]{dis03}
\usepackage{epsf,amsmath}

\begin{document}

\title{THE CONSEQUENCES OF THE RELATIONS BETWEEN \\
NON-SINGLET CONTRIBUTIONS TO $g_1^N$ AND $F_1^N$ \\
STRUCTURE FUNCTIONS WITHIN INFRARED RENORMALON MODEL
\thanks{This work is
supported by the RFBR Grants Nos.02-01-00601,03-02-17047
 and 03-02-17177}}

\author{A.~L.~KATAEV \\
Institute for  Nuclear Research
of the Academy of Sciences of Russia \\
Moscow 117312, Russia\\
E-mail: kataev@ms2.inr.ac.ru }

\maketitle

\begin{abstract}
\noindent 
We report on the following consequences of the relations between non-singlet 
contributions to $g_1^N$ and $F_1^N$ structure functions found  within 
infrared renormalon model:  the discovery of 
new  next-to-leading order inequalities between non-singlet 
polarized and unpolarized parton densities,  and the existence 
of the effects of similarity between non-perturbative and perturbative 
contributions to the unpolarized and polarized Bjorken sum rules. 
\end{abstract}

\section{Introduction}          
The infrared renormalon approach is the popular model for simulating the 
behavior of the high-twist corrections in the 
$\overline{MS}$-scheme  (see the reviews of 
Refs.\cite{Beneke:1998ui,Beneke:2000kc} an the recent application of 
Ref.\cite{Broadhurst:2000yc}). In particular, in Ref.\cite{Dasgupta:1996hh}
the prediction for the $x$-shape of the twist-4 contributions to the
non-singlet 
(NS) structure function (SF)  $xF_3$ and the  NS parts of SFs 
$g_1^N$ and $F_1^N$ 
were obtained. It should be stressed, 
that the qualitative validity of the IRR 
model for twist-4 term of $xF_3$ SF from Ref. \cite{Dasgupta:1996hh}
was confirmed  during the NLO fits 
of Refs.\cite{Kataev:1997nc} (for the brief discussions see the talks of 
Ref.\cite{Kataev:2002rj}). In this report we will use the machinery of the 
IRR model to find definite relations, induced by the IRR model predictions 
of Ref.\cite{Dasgupta:1996hh} for the $1/Q^2$ corrections to 
the  NS parts of $g_1^N$ and $F_1^N$ SFs
(see Ref.\cite{Kataev:2003jv}) and for the  Bjorken sum rules (both 
polarized  and unpolarized \cite{Broadhurst:2002bi}).  
  
\section{The bounds on the NS polarized parton densities}
Let us consider the expression for the asymmetry of photon-nucleon
polarized DIS, which in general has the following form
\begin{equation}
A_1^N(x,Q^2)=(1+\gamma^2)\frac{g_1^N(x,Q^2)}{F_1^N(x,Q^2)}+
\frac{h^{A_1}(x)}{Q^2}
\end{equation}
where $\gamma=4M_N^2x^2/Q^2$ and the second term in the r.h.s. of Eq. (1) 
is the dynamical $1/Q^2$ correction. It should be stressed, that 
the recent results of the fits to combined CERN, DESY and SLAC polarized DIS 
data in the kinematic region $0.005\leq x\leq 0.75$ and $1~{\rm GeV}^2\leq 
58~{\rm GeV^2}$ demonstrated, that the $x$-shape of $h^{A_1}(x)$ is consistent
with zero \cite{Leader:2001kh}. This effect is leading to the conclusion that 
\begin{equation}
\frac{h^{g_1}(x)}{Q^2g_1^N(x,Q^2)}\approx \frac{h^{F_1}(x)}{Q^2F_1^N(x,Q^2)}~~.
\end{equation}
Using now the well-known inequality 
\begin{equation}
|g_1^N(x,Q^2)|\leq F_1^N(x,Q^2)
\end{equation}
we get \cite{Kataev:2003jv}, that 
\begin{equation}
|h^{g_1}(x)|\leq |h^{F_1}(x)|~~~~.
\end{equation}
It should be stressed that the IRR model considerations 
of Ref.\cite{Dasgupta:1996hh} predict that in the NS approximation $h^{F_1}(x)$
and $h^{F_3}(x)$ have the same form, namely 
\begin{equation}
\label{F_1}
h^{F_1}(x,\mu^2)=h^{F_3}(x,\mu^2)=A_2^{'}\int_x^1\frac{dz}{z}C_1(z)
q^{NS}(x/z,\mu^2)
\end{equation}
where $C_1(z)=-4/(1+x)_{+}+2(2+x+2x^2)-5\delta(x)-\delta^{'}(1-x)$ is the 
calculated in Ref.\cite{Dasgupta:1996hh} IRR model coefficient function
and 
\begin{equation}
q^{NS}(x,\mu^2)=\sum_{i=1}^{n_f}\bigg(e_i^2-\frac{1}{n_f}\sum_{k=1}^{n_f}e_k^2\bigg)
(q_i(x,\mu^2)+\overline{q}(x,\mu^2))
\end{equation}
are the NS parton densities, $\mu^2$ is the normalization point of order 
$1~{\rm GeV}^2$ and $A_2^{'}$ is the IRR model parameter, which should be 
extracted from the fits of concrete data. 
It should be noted that the identity of Eq. (\ref{F_1}) does not contradict 
point of view that to study the $Q^2$ behavior of $A_1(Q^2)$ in the NS 
approximation it might be convenient to use the concrete $xF_3$ data instead 
of theoretical expression for $F_1^N$ \cite{Kotikov:1998ew}. 

In the case of NS approximation for the $g_1^N$ SF the IRR result of 
Ref.\cite{Dasgupta:1996hh} has the following form   
\begin{equation}
h^{g_1}(x,\mu^2)=A_2\int_x^1\frac{dz}{z}C_1(z)\Delta^{NS}(x/z,\mu^2)
\end{equation}
where the IRR model coefficient function has the same expression as in the 
case of the IRR model predictions for $h^{F_1}(x,\mu^2)$ and 
\begin{equation}
\Delta^{NS}(x,\mu^2)=\sum_{i=1}^{n_f}\bigg(e_i^2-\frac{1}{n_f}\sum_{k=1}^{n_f}
e_k^2
\bigg)
(\Delta q_i(x,\mu^2)+\Delta\overline{q}_i(x,\mu^2))
\end{equation} 
are the NS polarized parton densities. Combining now the NS  expressions for 
$h^{F_1}(x,\mu^2)$ and $h^{g_1}(x,\mu^2)$ we arrive to the main result 
of the work of Ref.\cite{Kataev:2003jv}, namely 
\begin{equation}
|A_2\Delta^{NS}(x,\mu^2)|\leq| A_2^{'}q^{NS}(x,\mu^2)|
\end{equation}
which is valid both at the LO and NLO. In the case of $|A_2|\sim |A_2^{,}|$
assumed in Ref.\cite{Dasgupta:1996hh}, this inequality is similar to the LO 
bound of Ref. \cite{Altarelli:1998gn},
namely 
\begin{equation}
|\Delta(x,Q^2)|\leq q(x,Q^2)
\end{equation}

\section{The relations between renormalon contributions 
to unpolarized and polarized Bjorken sum rules}
It should be stressed that the relations between renormalon contributions 
to NS parts of $g_1^N$ and $F_1^N$ SFs are also  manifesting themselves 
in the case of consideration of theoretical predictions for the 
unpolarized Bjorken sum rule 
\begin{equation}
C_{Bjunp}=\int_0^1dx\bigg[F_1^{\nu p}(x,Q^2)-F_1^{\nu n}(x,Q^2)\bigg]
\end{equation}
and polarized Bjorken sum rule 
\begin{equation}
\int_0^1dx\bigg[g_1^p(x,Q^2)-g_1^n(x,Q^2)\bigg]=\frac{1}{3}|\frac{g_A}{g_V}|
C_{Bjp}(Q^2)
\end{equation}
which were calculated in the large $N_F$-limit in Ref.\cite{Broadhurst:2002bi}
and in Ref. \cite{Broadhurst:1993ru} correspondingly.
Indeed, the large $N_F$-limit of the perturbative part of unpolarized 
Bjorken sum rule has the following form \cite{Broadhurst:2002bi}
\begin{eqnarray}
C_{Bjunp}&=&1+\frac{C_F}{T_FN_F}\sum_n^{\infty}U_n
\bigg(T_FN_F\overline{a}_s\bigg)^n+O(1/N_F^2) \\ \nonumber 
U_n&=&lim_{\delta\rightarrow 0}\bigg(-\frac{4}{3}\frac{d}{d\delta}\bigg)^{n-1}
U(\delta) \\ \nonumber
U(\delta)&=&-\frac{2exp(5\delta/3)}{(1-\delta)(1-\delta^2/4)}
\end{eqnarray}
The large $N_F$-expression for the Bjorken polarized sum rule can be obtained 
from the following equations  \cite{Broadhurst:1993ru}:
\begin{eqnarray}
C_{Bjp}&=&1+\frac{C_F}{T_FN_F}\sum_n^{\infty}K_n
\bigg(T_FN_F\overline{a}_s\bigg)^n+O(1/N_F^2) \\ \nonumber 
K_n&=&lim_{\delta\rightarrow 0}\bigg(-\frac{4}{3}\frac{d}{d\delta}\bigg)^{n-1}
K(\delta) \\ \nonumber
K(\delta)&=&-\frac{(3+\delta)exp(5\delta/3)}{(1-\delta^2)(1-\delta^2/4)}~~.
\end{eqnarray}
Notice that like in the case discussed in the previous 
Section, the renormalon contributions to unpolarized and polarized 
Bjorken sum rules are related as 
\begin{equation}
K(\delta)=\bigg(\frac{3+\delta}{2(1+\delta)}\bigg)U(\delta)~~~.
\end{equation}
This, in turn, results in the similarity  between twist-4 contributions 
to both sum rules and explains the similarity between perturbative 
theory predictions to both sum rules, observed in Ref. \cite{Gardi:1998rf}.
More detailed studies of the consequences of the results of this Section 
are on the agenda.


\begin{thebibliography}{0}

\bibitem{Beneke:1998ui}
M.~Beneke,
Phys.\ Rept.\  {\bf 317} (1999) 1
[hep-ph/9807443].
\bibitem{Beneke:2000kc}
M.~Beneke and V.~M.~Braun,
hep-ph/0010208.
\bibitem{Broadhurst:2000yc}
D.~J.~Broadhurst, A.~L.~Kataev and C.~J.~Maxwell,
Nucl.\ Phys.\ B {\bf 592} (2001) 247
[hep-ph/0007152].
\bibitem{Dasgupta:1996hh}
M.~Dasgupta and B.~R.~Webber,
Phys.\ Lett.\ B {\bf 382} (1996) 273
[hep-ph/9604388].
\bibitem{Kataev:1997nc}
A.~L.~Kataev, A.~V.~Kotikov, G.~Parente and A.~V.~Sidorov,
Phys.\ Lett.\ B {\bf 417} (1998) 374
[hep-ph/9706534];\\
S.~I.~Alekhin and A.~L.~Kataev,
Phys.\ Lett.\ B {\bf 452} (1999) 402
[hep-ph/9812348];\\
A.~L.~Kataev, G.~Parente and A.~V.~Sidorov,
Nucl.\ Phys.\ B {\bf 573} (2000) 405
[hep-ph/9905310];\\
A.~L.~Kataev, G.~Parente and A.~V.~Sidorov,
Phys.\ Part.\ Nucl.\  {\bf 34} (2003) 20
[Fiz.\ Elem.\ Chast.\ Atom.\ Yadra {\bf 34} (2003) 43]
[hep-ph/0106221].
\bibitem{Kataev:2002rj}
A.~L.~Kataev, G.~Parente and A.~V.~Sidorov,
Nucl.\ Phys.\ Proc.\ Suppl.\  {\bf 116} (2003) 105
[hep-ph/0211151];\\
A.~L.~Kataev, G.~Parente and A.~V.~Sidorov,
J. \ Phys. {\bf G} (2003) (in press),
hep-ph/0209024.
\bibitem{Kataev:2003jv}
A.~L.~Kataev,
JETP \ Lett. {\bf 77} (2003) 458,
[hep-ph/0302101].
\bibitem{Broadhurst:2002bi}
D.~J.~Broadhurst and A.~L.~Kataev,
Phys.\ Lett.\ B {\bf 544} (2002) 154
[hep-ph/0207261].
\bibitem{Leader:2001kh}
E.~Leader, A.~V.~Sidorov and D.~B.~Stamenov,
Eur.\ Phys.\ J.\ C {\bf 23} (2002) 479
[arXiv:hep-ph/0111267];\\
E.~Leader, A.~V.~Sidorov and D.~B.~Stamenov,
Phys.\ Rev.\ D {\bf 67} (2003) 074017
[hep-ph/0212085].
\bibitem{Kotikov:1998ew}
A.~V.~Kotikov and D.~V.~Peshekhonov,
Eur.\ Phys.\ J.\ C {\bf 9} (1999) 55
[hep-ph/9810224].
\bibitem{Altarelli:1998gn}
G.~Altarelli, S.~Forte and G.~Ridolfi,
Nucl.\ Phys.\ B {\bf 534} (1998) 277
[hep-ph/9806345].
\bibitem{Broadhurst:1993ru}
D.~J.~Broadhurst and A.~L.~Kataev,
Phys.\ Lett.\ B {\bf 315} (1993) 179
[hep-ph/9308274].
\bibitem{Gardi:1998rf}
E.~Gardi and M.~Karliner,
Nucl.\ Phys.\ B {\bf 529} (1998) 383
[hep-ph/9802218].

\end{thebibliography}
\end{document}